\documentstyle[12pt]{article}
\begin{document}

\author{Mario Castagnino \\
Instituto de Astronom\'{\i}a y F\'{\i}sica del Espacio.\\
Casilla de Correos 67, Sucursal 28.\\
1428 BUENOS AIRES, Argentina.}
\title{The Mathematical Structure of Superspace as a Consequence of Time 
Asymmetry}
\maketitle

\begin{abstract}
It is demonstrated how a convenient choice of the mathematical structure of
the quantum cosmology superspace, precisely the definition of a convenient
regular state superspace and the restriction of the dynamics to this space,
yields directly to an irreversible evolution, in the classical (and
semiclassical) phase of the universe, where:

\begin{itemize}
\item  Decoherence and correlations take place and therefore give origin to
a classical universe.

\item  The second law of thermodynamic is demonstrated.

\item  Connection with Reichenbach branch system idea can be implemented.

\item  Some rough coincidence with observational data are obtained.

\item  The arrows of time can be correlated.

\item  Time asymmetry can be explained as a state space asymmetry ( e. g.
like a spontaneous symmetry breaking)

All these facts solve the problem of time-asymmetry and show that it is time
asymmetry itself that defines the most important features of mathematical
structure of superspace.

\item  e-mail: Castagni (a) iafe.uba.ar

\item  Pacs Nrs. 05.20-y, 03.65 BZ, 05.30-d.
\end{itemize}
\end{abstract}

\section{Introduction}

The role of physics is to explain nature in the {\it best possible way.}
Therefore physicists use to consider a set of physical phenomena and to
choose the best axiomatic structure to mimic these phenomena. This axiomatic
structure contains a mathematical structure and a set of axioms (or
postulates, or principles, or hypotheses), stated using the language of the
chosen mathematical structure. If the mathematical structure is the most
naturally related to the set of phenomena and a minimal number of axioms is
used, physicist say that they have explained nature in the {\it best
possible way. }But, frequently a better mathematical structure and a smaller
set of axioms are found to explain a larger set of physical phenomena, then
physicists say that they understand the problem even better because, in
fact, they have found a better explanation {\it i. e. a more economical one. 
}But the chosen mathematical structure and the chosen axioms cannot be
explained by themselves, since the only motivation of the choice is to mimic
nature in the best possible way.

Thus, physical phenomena are not a consequence of the chosen mathematical
structure, quite on the contrary, the choice of the mathematical structure
is a consequence of the physical phenomena that we are trying to explain.

Gravitation was explained by Einstein choosing a riemannian manifold as
mathematical structure and postulating that particles space-time paths were
the geodesic of such a manifold, etc. The only explanation of this choice is
that the theoretical motions, so described, mimic real motions better than
the motions described using other curves or other kind of manifolds (flat
space-time, projective manifolds, etc.). But the riemannian manifold and the
geodesics cannot be explained by themselves. In fact, the choice of the
riemannian manifold, as the mathematical structure to explain gravity, is
really motivated by the method we use to measure time and distance \cite{JMP}%
.

It also happens that, when a new or unfamiliar mathematical structure is
introduced, some physicists think that the new structure is introduced {\it %
by hand}, because they do not realize that {\it every} mathematical
structure was introduced by hand in, order to explain nature in the best
possible way.

In the perspective of this pedagogical (perhaps pedantic but necessary)
introduction we will choose:

\subsection{The set of phenomena.}

The set of phenomena considered in this paper will be those of usual quantum
cosmology (QC) based in Wheeler-De Witt (WDW) equation \cite{DeWitt}, \cite
{Halliwell}, plus those of irreversible statistical quantum physics, like
the definitions of the various arrows of time, the final equilibrium state,
decoherence, correlations, etc. (see the book \cite{Halliwellbook} where
almost all subjects, results, and problems can be found).

\subsection{The mathematical structure.}

QC is based in WDW equation; 
\begin{equation}
H\Psi =0  \label{T.1.1}
\end{equation}
where $H$ is the hamiltonian operator of the model for the considered
universe and $\Psi $ the wave function of the universe (that will be also
called $|\Psi >)$ \cite{Halliwell}, \cite{Hartle}. Usually $H$ is well
defined by the model we are studying, but the real mathematical nature of $%
\Psi $ is not so well defined. In fact, let us call ''$q"$ the configuration
variables and $Q$ the configuration space, then $\Psi =\Psi (q)$ is a
function of the configuration variables. We must define the space of wave
functions $\Psi .$ Let us suppose that this space has a discrete basis $%
{\cal P}$ of certain functions of the $q$ (let us say polynomials of the $q$
or polynomials multiplied by convenient dumping factors for those $q,s$ that
goes to infinity). But, this set is not enough because usually wave
functions belong to a complete space. Then we must complete the space
expanded by ${\cal P}$ with some topology. The usual idea is to complete the
space expanded by ${\cal P}$ with the topology of the norm. Even if the
choice of a norm contains a very important physical problem: the definition
of probability, we are now exclusively interested in the mathematical
problem of how to complete ${\cal P}$ with an adequate topology. Then we
have really two problems:

i.- We must define a norm in the space expanded by ${\cal P}$ something like 
\cite{Hawking}: 
\begin{equation}
\parallel \Psi \parallel =\int_Q\Psi ^{*}(q)\Psi (q)dq  \label{T.1.2}
\end{equation}
But we can also foliate $Q$ with hypersurfaces $\Sigma $ and define \cite
{DeWitt}: 
\begin{equation}
\parallel \Psi \parallel =\int_\Sigma \Psi^{*}(q)
\partial_n\Psi (q)d\sigma _n  \label{T.1.3}
\end{equation}
where $n$ is the normal to $\Sigma $ and $d\sigma _n$ is the hypervolume of
the hypersurface $\Sigma $ differential element (this quantity will be
positive definite in a convenient, subspace of superspace). Or we can
consider that really $\Psi $ is an operator and then we must go to a third
quantization \cite{Tcachev}, etc. So we have many possibilities to define
the norm. Actually we are trying to copy usual quantum mechanics and the
essential property to define an adequate norm is that {\it it would be a
constant under time evolution.} But in QC there is no time \cite{Kuchar}. 
\cite{Barbour} \cite{Lumen}. In fact, eq. (\ref{T.1.1}) is not a
Schroedinger equation but an eigen equation that defines a ''stationary''
eigenfunction, with no time variable in it. Therefore, as the definition of
norm is intimately intertwined with the problem of the definition of time in
QC, and this problem is not solved \cite{Kuchar}, \cite{Probtime}, \cite
{Isham} it is not possible to give a definitive definition of norm. Really
in this paper we will adopt the conservative attitude that this problem
cannot be solved and therefore QC is a timeless theory \cite{Kuchar}, \cite
{Barbour}. As a consequence it is very difficult to define a satisfactory
norm.

ii.-But even if a norm would be chosen, say (\ref{T.1.2}), we have another
problem. If we complete the space expanded by ${\cal P}$ with (\ref{T.1.2})
we will find ${\cal H=}L^2(Q)$ the Hilbert space of square integrable
functions over $Q.$ But $H$ is an operator with derivatives and the
functions of ${\cal H}$ are, in general, not derivable, they are just square
integrable, so for an arbitrary function $\Psi $ of ${\cal H,}$ $H\Psi $ has
no meaning (and we would have the same problem with the other norms). So we
must choose another topology to complete the space expanded by ${\cal P}$.
Let us consider the Schwarz class functions, namely functions that can be
derived an infinite number of times and that are well behaved in the
eventual infinite of coordinates $q$ (precisely they vanish faster than any
polynomial). The set of Schwarz functions is a nuclear space. Then if we
complete the space expanded by ${\cal P}$ with the corresponding topology,
which is not a norm topology but a nuclear one \cite{Bogolyubov}, we obtain
a space ${\cal S}$ where we can derive any number of times. If ${\cal %
S^{\times }}$ is the space of (anti)linear functional over ${\cal S,}$ our
mathematical structure is really the Gel'fand triplet (or rigged Hilbert
space): 
\begin{equation}
{\cal S\subset H\subset S^{\times }}  \label{T.1.4}
\end{equation}
Then we can say that ${\cal S}$ is the superspace of {\it regular states}
where we can do all our computations and ${\cal S^{\times }}$ is the
superspace of {\it generalized states}, where we will find usual
distributions like Dirac's deltas or plane (or curved) waves, that can be
used to expand regular states. In ordinary quantum mechanics ${\cal H}$
would be the superspace of {\it states, }but in QC this space looses almost
all its importance because we do not know what norm we have to use for sure
(even if, to compute probabilities , we will use norm (\ref{T.1.2}) in spite
of the fact that we have other possible choices\cite{Pajaro}), while ${\cal S}$ and ${\times}$ are well defined, since their
definitions are norm-independent. The ''inner product'' $<F|\Psi >$ of an
element of ${\cal S^{\times }\ }$written as a ''bra'', $<F|,$ with an
element of ${\cal S}$ written like a ''ket'', $|\Psi >,$ is also well
defined, since it is just the functional acting over a regular state, (also
the product in the inverted order is well defined, if we set $<\Psi
|F>=<F|\Psi >^{*})$. So ${\cal S}$ is the arena where timeless QC works:
where time and norm have disappeared from the mathematical structure and are
substituted by the nuclear topology of ${\cal S}$.

But if we use ${\cal S}$ as a regular space we cannot encompass (in the most
economical way) irreversible statistical quantum mechanics, which has an
asymmetry that it is not contained in ${\cal S}$. There are only two causes
for asymmetry in nature: either the laws of nature are asymmetric or the
solutions of the equation of the theory are asymmetric. E. g.: The laws of
nature are asymmetric in the case of the weak interaction. The solutions of
the theory are asymmetric in the case of spontaneous symmetry breaking.

Time asymmetry is not an exception. Thus, if we want to retain the
time-symmetric laws of nature (namely the symmetry of eq. (\ref{T.1.1})) the
only way we have to explain the time-asymmetry of the universe or its
subsystems is to postulate that the space of solutions is not
time-symmetric, namely we use the second cause of asymmetry. So the proper
way to solve the problem is simply to define a realistic time-asymmetric
space of physical admissible solutions $\phi _{-},$ i. e. an adequate
mathematical structure for superspace. $\phi _{-}$ will contain the states
that evolve in an admissible way (e.g.: Gibbs ink drop spreading in a glass
of water, a sugar lump solving in a cup of coffee, etc.) and will not
contain the non admissible evolutions ( the ink or the sugar concentrating
spontaneously and creating the drop or the lump). The problem is to choose $%
\phi _{-}$ in the{\it \ best possible way. }As we will see{\it \ }we will
choose one with the required asymmetry. We will follow a heuristic approach:
we will suppose that $H$ is endowed with all the properties necessary to
define a reasonable universe. Obviously these ''realistic hamiltonians'' are
the only $H$ that we must consider. Therefore practically, in $H$ there must
be always some fields, like the mater field, the electromagnetical field,
and also the gravitational field (namely only the graviton field). So we
would write this hamiltonian in a midi-superspace way as: 
\begin{equation}
H=h_{nf}(g_j,\pi j)+h_f(\varphi ,p_\varphi )+h_i(g_{ij,}\varphi )
\label{T.1.5}
\end{equation}
where $h_{nf}(g_{j,}\pi _j)$ is the ''no field'' hamiltonian (a function of
a discrete set of modes of the gravitational field $g_j$ and the
corresponding momenta $\pi _j)$, $h_f$ is the ''fields'' hamiltonian, (let
us say the continuous set of all the modes of just one field $\varphi $ ,
that represents all the physical fields in our model, and the corresponding
momenta $p_\varphi )$, and $h_i$ is the interaction hamiltonian among fields
and no fields (usually an interaction among only the configuration
variables).

To complete the panorama, (using the Gel'fand-Maurin theorem \cite{Chiu}) we
can find, for the partial hamiltonian $h_{nf}+h_f$, a spectral decomposition:

\begin{equation}
h_{nf}+h_f=\sum_i\omega _i|i><i|+\int_0^\infty \omega |\omega ><\omega
|d\omega  \label{T.1.8}
\end{equation}
where the first term of the r.h.s. corresponds to the spectral decomposition
of $h_{nf},$ (cf. eq. \ref{T.1.5}) and therefore to a discrete spectrum, and
the second one to the spectral decomposition of $h_f$ and therefore to a
continuous spectrum. Some negative eigenvalues ''$\omega "$ must appear
because $H$ is not bounded from below. Any state $|\Psi >\in {\cal S}$ can
be expanded as: 
\begin{equation}
|\Psi >=\sum_i|i><i|\Psi >+\int_0^\infty |\omega ><\omega |\Psi >d\omega
\label{T.1.9}
\end{equation}
This is the spectral decomposition of $|\Psi >$ in basis $\{|i>,|\omega >\}$%
. In order that $|\Psi >$ would satisfy the WDW eq. (\ref{T.1.1}) it must
be: 
\begin{equation}
H|\Psi >=\sum_i\omega _i|i><i|\Psi >+\int_0^\infty \omega |\omega ><\omega
|\Psi >d\omega +h_i|\Psi >=0  \label{T.1.10}
\end{equation}
Most likely this equation can be solved.

Also, in every reasonable model, the universe will end in a classical phase,
going first through semiclassical one. Namely, a variable $a$ will exist
(one of the $g_j$ or a function of the $g_j$ ) such that a time $\eta $ can
be defined as a function of $a$. When $\eta \rightarrow \infty $ we will
obtain a classical geometry $g_{\mu \nu }^{out}$ for the universe. It will
be the most probable geometry of the universe (i. e. the geometry that
appears most frequently) \cite{Atractor}. Using time $\eta $ we can
transform eq. (\ref{T.1.1}) in a Schroedinger equation, with the
corresponding hamiltonian $h$ \cite{Hartle}. Then using $h$ and the
classical geometry $g_{\mu \nu }^{out}$ we can find a semi-classical vacuum
state $|0,out>$ for the fields (the so called adiabatic vacuum), that
diagonalizes the hamiltonian $h=h(g_{\mu \nu }^{out})=h(out)$ (computed in
the geometry $g_{\mu \upsilon }^{out}),$ the creation and annihilation
operators related to this vacuum, and the corresponding Fock space. But the
only essential ingredient we need to implement the theory is the $h(out).$
Then, using these objects, we can find a set of eigenvectors $|\omega ,out>$ 
\cite{SE}, such that: 
\begin{equation}
h|\omega ,out>=\omega |\omega ,out>  \label{T.1.6}
\end{equation}
where $\omega $ is a continuous eigenvalue of $h_{}$ (say $0\leq \omega
<\infty )$ \cite{Cont}. So: 
\begin{equation}
h=\int_0^\infty \omega |\omega ,out><\omega ,ou|d\omega  \label{T.1.7}
\end{equation}
where $|\omega ,out>\in {\cal S^{\times },\ }$thus if $|\Psi >\in {\cal S,\ }%
<\omega ,out|\Psi >$ is well defined and so $h|\Psi >$. The existence of
this kind of expansions can be also considered a consequence of
Gel'fand-Maurin theorem \cite{Chiu}. But all these manipulations are just
formal, so we will be sure that what we are doing is correct only in
concrete examples, as the one in the next section. In fact, we will find all
this mathematical elements in the model presented there. Surely we will also
find these elements in more complex models \cite{Necessary}.

Now, let us define our new regular state space $\phi _{-}\subset {\cal S}$.
Precisely, we can promote $\omega $ to a complex variable $z$ and ask, not
only that $|\Psi >\in {\cal S,}$ but also that $<z,out|\Psi >$ would be an
analytic function in the lower complex halfplane (precisely that $<\omega
,out|\Psi >\in H_{-}^2$ being $H_{-}^2$ the Hardy functions class from
below). These functions $|\Psi >$ belong to a space $\phi _{-}$ such that: 
\begin{equation}
\phi _{-}\subset {\cal S}  \label{T.1.11}
\end{equation}
and we have a new Gel'fand triplet: 
\begin{equation}
\phi _{-}\subset {\cal H}\subset \phi _{-}^{\times }  \label{T.1.12}
\end{equation}
then we know that \cite{Bogolyubov}: 
\begin{equation}
{\cal S}^{\times }\subset \phi _{-}^{\times }  \label{T.1.12'}
\end{equation}
So we have restricted the regular state superspace and simultaneously we
have enlarged the generalized state superspace so we will have more general
spectral expansions (this fact will be of outmost importance). But, as we
can as well choose the upper complex halfplane (precisely $<\omega ,out|\Psi
>\in H_{+}^2$ being $H_{+}^2$ the Hardy functions class from above), we also
have another space $\phi _{+}$ such that: 
\begin{equation}
\phi _{+}\subset {\cal S}  \label{T.1.13}
\end{equation}
and another Gel'fand triplet: 
\begin{equation}
\phi _{+}\subset {\cal H}\subset \phi _{+}^{\times }  \label{T.1.14}
\end{equation}
We also know that \cite{Bogolyubov}: 
\begin{equation}
{\cal S}^{\times }\subset \phi _{+}^{\times }  \label{T.1.14'}
\end{equation}
Now we can also say that we have obtained the space $\phi _{-}$ completing
the space expanded by ${\cal P}$ with the nuclear topology of $\phi _{-}$,
namely the ${\cal S}$-topology restricted to $\phi _{-}$ (and the same thing
can be said about $\phi _{+}).$ Clearly this topology is endowed with a new
asymmetry which ${\cal S}$ does not have. Precisely, this asymmetry allows
us to choose between $\phi _{-}$ or $\phi _{+}$ even if we maintain all the
symmetries of $H$ \cite{Lombardo?}. Thus we can break one of these
symmetries, restricting the dynamics to superspace $\phi _{-}$ which then
would be considered as the superspace of regular states. As we will see this
restriction produces the desired time asymmetry. Frequently physicists make
analytic continuation in the complex energy plane supposing that some
functions are analytic in one halfplane only. In these cases they are
implicitly using the kind of mathematical structure we have explicitly
introduced here, so the idea is, by no means, new.

Thus our mathematical structure will essentially be (\ref{T.1.12}), $\phi
_{-}$ will be our superspace of regular states, where we must find the
states that satisfy the WDW eq. (\ref{T.1.1}) and $\phi _{-}^{\times }$ will
be our generalized state superspace. From (\ref{T.1.11}) we see that we have
restricted our regular state superspace, so nothing unphysical can happen.
We are just adding a new requirement to regular states, in order to assure
their asymmetry. Furthermore $\phi _{-}$ is dense in ${\cal H}$, so if
someone would believe that the real ''physical'' states belong to ${\cal H}$
we could argue that these ''physical'' states can be approximated, by
regular states of $\phi _{-},$ as close as we want.

\subsection{The axiomatic structure.}

We do not pretend to give a completely rigorous axiomatic structure in this
paper (but just an approximation of it). Furthermore we do not know if the
proposed axiomatic structure is unique. We are just proposing a first draft
of a complete axiomatic structure, so we will call to our axioms just
hypotheses.

As we would like that the equilibrium state would be contained in our theory
we must also consider mixed states $\rho $ and, therefore, the spaces: 
\begin{equation}
\Phi _{-}=\phi _{-}\otimes \phi _{-},....{\cal L=H\otimes H},....\Phi
_{-}^{\times }=\phi _{-}^{\times }\otimes \phi _{-}^{\times }  \label{T.1.15}
\end{equation}
and work in the Liouville triplet: 
\begin{equation}
\Phi _{-}\subset {\cal L}\subset \Phi _{-}^{\times }  \label{T.1.16}
\end{equation}
where ${\cal L}$ is the usual Liouville space of ordinary mixed states,
which actually we will never use, since our regular state superspace is $%
\Phi _{-.}$ So, let $\rho $ be a selfadjoint density matrix. Then our main
hypotheses are{\it :}

{\it H}$_1$.{\it \ The state $\rho $} {\it of the universe satisfies the
equations \cite{WdW}: 
\begin{equation}
H\rho =0  \label{T.1.18}
\end{equation}
}

{\it H$_2$}. {\it The state $\rho $} {\it of the universe belongs to the
superspace $\Phi _{-}$}, i. e.{\it : 
\begin{equation}
\rho \in \Phi _{-}  \label{T.1.19}
\end{equation}
}

{\it H$_3.$} $\rho (q,q^{\prime })$ {\it is proportional to the correlation
between the configurations }$q$ and $q^{\prime }$ {\it and }$\rho (q,q)$ 
{\it is proportional to the probability to find the configuration }$q$ {\it %
in the universe.}

These three axioms correspond to the three elements necessary to go into QC
mentioned in the introduction of paper \cite{Halliwell}: {\it dynamics,
''initial condition'' (precisely definition of the physically admissible
states of the universe in a timeless formalism), and interpretation.} Of
course H$_2$ alone does not fix the actual state of the universe, but any
state of the universe we choose to build our theory must be contained in $%
\Phi _{-},$ if we want that time-asymmetry would appear in a natural way. In
this paper we do not address the problem to find the real and unique state
of the universe, but only to define a superspace of admissible states such
that the universe would turns out to be time-asymmetric.

We will see how far we can go with this axiomatic structure. The paper is
organized as follows:

\begin{itemize}
\item  In section 2 we introduce our model, its semiclassical approximation
and we obtain a new spectral decomposition, using the regular superspace of
hypothesis H$_2.$

\item  In section 3 we obtain the evolution equation of the states.

\item  In section 4 up to section 9 we find the physical characteristics of
the model.

\item  In section 10 we draw our main conclusions.

\item  Two appendices, containing further explanations and the way to fix
the two $\pm $ signs of the theory, complete the paper.
\end{itemize}

\section{The Model.}

Let us see how we can implement all we have said in a simple model.

Let us consider the model of section 3, of ref. \cite{Paz}, or better the
one of ref. \cite{Lombardo}, where a Robertson-Walker metric: 
\begin{equation}
ds^2=a^2(\eta )(d\eta ^2-dx^2-dy^2-dz^2)  \label{T.2.0}
\end{equation}
is studied, (we will mostly consider the flat space geometry case). The
total action is: $S=S_g+S_f+S_i,$ being $S_g(=S_{nf})$ the gravitational
action, $S_f$ the usual action of an spinless massless field $\varphi ,$
conformally ($\xi =\frac 16)$ coupled , and $S_i$ the interaction given by a
mass term in Robertson-Walker geometry. The gravitational action is given
by: 
\begin{equation}
S_g=M^2\int d\eta [-\frac 12\stackrel{\bullet }{a}^2-V(a)]  \label{T.2.1}
\end{equation}
where $M$ is the Planck's mass, $\eta $ is the conformal time, $a$ is the
Robertson-Walker scale , $\stackrel{\bullet }{a}=\frac{da}{d\eta },$ and $%
V(a)$ is a potential that arises from the spatial curvature, a possible
cosmological constant, and, eventually a classical matter field. $V(a)$ is a
potential with a bounded support contained in $0\leq a\leq a_1$, with $%
a_1\gg 0$, (in many examples $V(a)$ is a function of $a^2$ and $V(a)$
strongly vanishes, for $a^2\rightarrow \infty ,$ \cite{Lombardo} so our
potential can be considered as a good approximation of these examples). This
case is the simplest of all, but we believe that the main features, that we
will find, will also be present in more general cases.

The WDW eq. (\ref{T.1.1}) for our model is: 
\begin{equation}
H\Psi (a,\varphi )=(h_g+h_f+h_i)\Psi (a,\varphi )=0  \label{T.2.2}
\end{equation}
where (in our the flat space geometry case): 
\begin{equation}
h_g=\frac 1{2M^2}\partial _a^2+M^2V(a)  \label{T.2.3}
\end{equation}
\begin{equation}
h_f=-\frac 12\int_k(\partial _{\varphi _k}^2-k^2\varphi _k^2)d{\bf k}
\label{T.2.4}
\end{equation}
\begin{equation}
h_i=\frac{m^2a^2}2\int_k\varphi _k^2d{\bf k}  \label{T.2.5}
\end{equation}
where $m$ is the mass of the scalar field, and $k^2=|{\bf k|^2}$, where $%
\frac{{\bf k}}a$ is the linear momentum of the field, in the flat case we
are working (in the two other cases, namely open and close space geometry,
the integrals of eqs. (\ref{T.2.4}) and (\ref{T.2.5}) are integrations on
adapted coordinates, in the open case, and summatories, in the close case,
where ${\bf k\ }$ is substituted by a discrete variable; see the
corresponding equations in \cite{Paz}).

Now, let us go to the semiclassical case using the WKB method \cite
{Halliwell}, \cite{Hartle}, \cite{DBohm}. So let: 
\begin{equation}
\Psi (a,\varphi )=\exp [iMS(a)]\chi (a,\varphi )  \label{WKB}
\end{equation}
To satisfy WDW eq. (\ref{T.1.1}), at order $M^2,$ the principal Jacobi
function $S(a)$ must satisfy the Hamilton-Jacobi equation: 
\begin{equation}
\left( \frac{dS}{da}\right) ^2=2V(a)  \label{T.2.6}
\end{equation}
Now we can define the time, in our up to now, timeless theory. It is the
(semi)classical time parameter $\eta =\eta (a)$ given by: 
\begin{equation}
\frac d{d\eta }=\frac{dS}{da}\frac d{da}=\pm \sqrt{2V(a)}\frac d{da}
\label{T.2.7}
\end{equation}
From eqs. (\ref{T.2.6}) and (\ref{T.2.7}) we can find the set of classical
solutions: 
\begin{equation}
a=\pm f(\eta ,C)  \label{T.2.8}
\end{equation}
where $C$ is an arbitrary integration constant. Using different values for
this constant and different choices of the $\pm $ sign we obtain different
classical geometries (in more general cases many constants would be
necessary). For $a>a_{1,}$ it is $\sqrt{2V(a)}=0$ (since $V(a)$ have a
bounded support, contained in $[0,a_1]),$ and we cannot define time $\eta $
using eq. (\ref{T.2.7}), thus we must choose another hand for our clock to
define the time there. To avoid this problem let us consider that when $%
a>a_1 $ it is $\sqrt{2V(a)}=\varepsilon =const.\cong 0.$ We can always make $%
\varepsilon =0$ to reobtain. the real case. Then $a$ will be 
\begin{equation}
a=\pm \varepsilon \eta +C  \label{T.2.9}
\end{equation}
So we can see that the potential can also be considered as a function with
bounded support in variable$\eta .$ We will consider that $\varepsilon >0$.
The role of $C$ is just to fix the origin of time, so we can take any $C$ we
want . As the coupling is conformal we will have well defined vacua \cite
{CastagninoQFTCST}, \cite{Birrell}. In particular we can consider two scales 
$a_{in}$ and $a_{out}$ such that $0<a_{in}\ll a_1,$ $a_{out}\gg a_1$ and
define the $|0,in>,|0,out>$ vacua there. (We can as well transform all the
equations to the non-rescaled case, consider the proper time $t=\int ad\eta
, $ and the physical momentum $\frac{{\bf k}}a$ and define the $|0,out>$ in
the $\eta \rightarrow \infty $ limit, as in the appendix A of paper \cite
{Lombardo}, but here we will use the first simpler formalism.)

In this model we have: 
\begin{equation}
h(a)=h_f(\varphi _k)+h_i(a,\varphi _k)  \label{T.2.10}
\end{equation}
where we have omitted the $\varphi _k$ in $h(a).$ Then: 
\begin{equation}
h(a)=\frac 12\int_k[-\frac{\partial ^2}{\partial \varphi _k^2}+\Omega
_k^2(a)\varphi _k^2]d{\bf k}  \label{T.2.11}
\end{equation}
where (cf. eqs. (\ref{T.2.4}) and (\ref{T.2.5})): 
\begin{equation}
\Omega _k^2(a)=m^2a^2+k^2  \label{T.2.12}
\end{equation}
So $h(a)$ is a time dependent hamiltonian, where all its time dependence
comes from a scale variable mass $m^2a^2.$ It is well known (\cite
{CastagninoQFTCST}, \cite{Birrell}) that we can diagonalize this time
dependent hamiltonian at $a_{in}$ and at $a_{out}$ and define the
corresponding vacua, the corresponding creation and annihilation operator,
and the corresponding Fock spaces. For the out-geometry the vacuum will be
the adiabatic vacuum, since $a_{out}\gg a_1,$ therefore all the out elements
will coincide with those defined in the introduction. In fact, the
out-geometry is almost constant during the final time evolution (that goes
up to $\eta \rightarrow \infty $) and therefore they correspond to the
geometry with the maximum probability$.$ $h(a_{out})$ reads: 
\begin{equation}
h(a_{out})=\int_k\Omega _k(a_{out}){\it a}_{out,k}^{\dagger }{\it a}_{out,k}d%
{\bf k}  \label{T.2.13}
\end{equation}
where ${\it a}_{out,k}^{\dagger }$ and ${\it a}_{out,k}$ are the creation
and annihilation operator corresponding to the out vacuum. With these
objects we can construct the Fock space with a basis: 
\begin{equation}
|k_1.k_2,...k_{n,out}>=|\{k\},out>\sim {\it a}_{out,k_1}^{\dagger }{\it a}%
_{out,k_2}^{\dagger }...{\it a}_{out,k_n}^{\dagger }|0,out>  \label{T.2.14}
\end{equation}
where we have called $\{k\}$ the set $k_1,k_2,...k_n.$ These states are
eigenvectors of $h(a_{out}),$ precisely; 
\begin{equation}
h(a_{out})|\{k\},out>=\omega (a_{out})|\{k\},out>  \label{T.2.15}
\end{equation}
(of course, this equation corresponds to eq. (\ref{T.1.6})) where: 
\begin{equation}
\omega (a_{out})=\sum_{k\in \{k\}}\Omega _k(out)  \label{T.2.16}
\end{equation}
We can use this energy to label the eigenvectors as: 
\begin{equation}
|\{k\},out>=|\omega ,[k],out>  \label{T.2.17}
\end{equation}
where $[k]$ is the remaining set of labels necessary to define the vector
unambiguously. $\{|\omega ,[k],out>\}$ is an orthonormal basis, namely: 
\begin{equation}
<\omega ,[k],out|\omega ^{\prime }[k^{\prime }],out>=\delta (\omega -\omega
^{\prime })\delta ([k]-[k^{\prime }])  \label{T.2.18}
\end{equation}
\begin{equation}
1=\int_0^\infty d\omega \int_{[k]}|\omega ,[k],out><\omega ,[k],out|d[k]
\label{T.2.19}
\end{equation}
where the meaning of the symbols related with $[k]$ is evident. In the close
space geometry case the indices would be discrete and the integral a
summatory.

All the same can also be done at $a_{in}.$ We can now define the S-matrix
between the in and out states: 
\begin{equation}
S_{\omega ,[k];\omega ^{\prime },[k^{\prime }]}=<\omega ,[k],in|\omega
^{\prime },[k^{\prime }],out>  \label{T.2.20}
\end{equation}
According to papers \cite{Audrecht}, \cite{Nusenbein}, this matrix has an
infinite set of complex poles as we will demonstrate in section 4 (also an
example is given in \cite{Lombardo} and using this paper and \cite{Birrell}
and \cite{Taylor} other examples can be obtained).

If we forget the indices $[k]$ and consider again the eq. (\ref{T.1.6}), we
see that $|\omega ,[k],out>$ is the $|\omega ,out>$ of this equation. In the
introduction we have defined the triplets (\ref{T.1.12}) and (\ref{T.1.14})
only using the hamiltonian $h(out)=h(a_{out}).$ These triplets correspond to
a Fock space defined for $a_{out}.$ But there will also be two similar
triplets defined in the Fock space at $a_{in}.$ We make the following choice
(motivated on reasons that will be apparent in a moment): for the in-Fock
space we will use functions $|\varphi >\in \phi _{+in,}$ namely such that $|$%
$\varphi >\in {\cal S}$ and $<\omega ,in|\varphi >\in H_{+}^2,$ and for the
out-Fock space we will use functions such that $|\varphi >\in \phi _{-,out},$
namely such that $|\varphi >\in {\cal S}$ and $<\omega ,out|\varphi >\in
H_{-}^2.$ So the $\phi _{-}$ of the introduction is now $\phi _{-,out}$ and
our regular states belong to this space \cite{S}. The role of $\phi _{+,in}$
is to allow us to define the corresponding functional space $\phi
_{+,in}^{\times }$ so we can use the functionals of this space in some
spectral decompositions \cite{Modifications}. As both vacua, at $a_{in},$
and $a_{out}$ are well defined and the particle production between these
vacua is finite and the theory is implementable \cite{Weder}. We can then
multiply the state of both Fock spaces.

So let us again write eq. (\ref{T.2.19}) with no $[k]:$%
\begin{equation}
1=\int_0^\infty d\omega |\omega ,out><\omega ,out|  \label{T.Z.1}
\end{equation}
Of course there is an analogous equation for the ''in'' case. Now using this
equation and eq. (\ref{T.2.20}) we have: 
\begin{equation}
|\omega ,out>=\int_0^\infty d\omega ^{\prime }|\omega ^{\prime },in><\omega
^{\prime }.in|\omega ,out>=\int_0^\infty d\omega ^{\prime }|\omega ^{\prime
},in>S_{\omega ^{\prime }.\omega }  \label{T.Z.2}
\end{equation}
Then: 
\begin{equation}
1=\int_0^\infty d\omega \int_0^\infty d\omega |\omega ^{\prime
},in>S_{\omega ^{\prime }.\omega }<\omega ,out|  \label{T.Z.3}
\end{equation}
or: 
\begin{equation}
<\psi |\varphi >=\int_0^\infty d\omega \int_0^\infty d\omega ^{\prime }<\psi
|\omega ^{\prime },in>S_{\omega ^{\prime },\omega }<\omega ,out|\varphi >
\label{T.Z.4}
\end{equation}
Now let $|\varphi >\in \phi _{-,out}$ and $|\psi >\in \phi _{+,in}.$ Then $%
<z,out|\varphi >\in H_{-}^2,<z,in|\psi >\in H_{+}^2$ and therefore $<\psi
|z,in>\in H_{-}^2.$ So in the integrand of the last equation all the factors
are analytic in the lower halfplane, with the exception of $S_{\omega
,\omega ^{\prime },}$ which has an infinite number of poles $\Omega _n$ as
we have already said. Then we can choose any curve $\Gamma ,$ beginning at
the origin, and going below all the poles of the lower halfplane up to the
infinity of the positive real axis \cite{A}. We can now change the
integration contour of eq. (\ref{T.Z.4}) from $[0,\infty )$ to the curve $%
\Gamma $ . If we add the pole contribution as in papers \cite{CGG}, \cite
{Bohm}, and \cite{Sudarshan}, we obtain: 
\[
<\psi |\varphi >=\sum_n<\psi |n+><n-|\varphi >+\int_\Gamma dz\int_\Gamma
dz^{\prime }<\psi |z^{\prime },in>S_{z^{\prime },z}<z,out|\varphi >= 
\]
\begin{equation}
=\sum_n<\psi |n+><n-|\varphi >+\int_\Gamma dz<\psi |z+><z-|\varphi >
\label{T.Z.5}
\end{equation}
where the summatory comes from the residues of the poles (each pole $\Omega_{n}$ is labelled by a discrete index $n$ of course $Im\Omega
_n\leq 0$) and: 
\begin{equation}
|z,->=|z,out>,......|z+>=\int_\Gamma dz^{\prime }S_{z^{\prime }z}|z^{\prime
},in>  \label{T.Z.6}
\end{equation}
Then, in a weak sense, we have found a new spectral decomposition of $1$: 
\begin{equation}
1=\sum_n|n+><n-|+\int_\Gamma dz|z+><z-|  \label{T.Z.7}
\end{equation}
Following the same procedure with $<\psi |h(out)|\varphi >$ we can obtain
the spectral decomposition of $h(out)$ (always in a weak sense): 
\begin{equation}
h(out)=\sum_n\Omega _n|n+><n-|+\int_\Gamma z|z+><z-|dz  \label{T.Z.8}
\end{equation}
We have three possibilities to choose the curve $\Gamma $:

i.- To use all possible curves $\Gamma $ as in paper \cite{Sudarshan}.

ii.- To take the curve (-$\infty ,0],$ in the second sheet, as in ref. \cite
{Bohm}, provided we have a good behavior at infinity of the lower halfplane.

iii.- To use the Nakanishi trick \cite{Nakanishi}, as in paper \cite
{Antoniou}, namely to define tilded functionals such that: 
\begin{equation}
\int_\Gamma |z+><z-|dz=\int_0^\infty |\widetilde{\omega +}><\widetilde{%
\omega -}|d\omega  \label{Naka}
\end{equation}
We will use this last notation. Then we have: 
\begin{equation}
1=\sum_n|n+><n-|+\int_0^\infty d\omega |\widetilde{\omega +}><\widetilde{%
\omega -|}  \label{T.Z.9}
\end{equation}
\begin{equation}
h(out)=\sum_n\Omega _n|n+><n-|+\int_0^\infty \omega |\widetilde{\omega +}><%
\widetilde{\omega -}|d\omega  \label{T.Z.10}
\end{equation}
(see also \cite{Trio} and paper \cite{CGG}, where the Nakanishi trick is
explain). From its own definition it is evident that $|n->,|z->,|\widetilde{%
\omega -}>\in \phi _{-,out}^{\times },$ since these vectors are functionals
over $|\varphi >\in \phi _{-,out}$ and that $|n+>,|z+>,|\widetilde{\omega +}%
>\in \phi _{+,in}^{\times }$ since these vectors are functionals over $|\psi
>\in \phi _{+,in}.$ Restoring the $[k]$ and eliminating the tilde (as we
will always do below) the last equation reads: 
\begin{equation}
h(out)=\sum_n\Omega _n|n+><n-|+\int_0^\infty d\omega \int_k\Omega _k|\omega
,[k]+><\omega ,[k]-|d[k]  \label{T.2.23}
\end{equation}
but we will continue with the previous shorthand notation and we will not
write the $[k]$ any more$.$ It can be proved that the bases $\{|n+>,|\omega
+>\},\{|n->,|\omega ->\}$ are a biorthonormal system, \cite{Bohm}, \cite
{Sudarshan}, and \cite{Antoniou} namely: 
\[
<n-|n^{\prime }+>=\delta _{nn^{\prime }},............<n-|\omega +>=0 
\]
\begin{equation}
<\omega -|n+>=0,...<\omega -|\omega ^{\prime }+>=\delta (\omega -\omega
^{\prime })  \label{T.2.27}
\end{equation}

From all these equation we have that: 
\[
h(out)|n+>=\Omega _n|n+> 
\]
\begin{equation}
<n-|h(out)=\Omega _n<n-|  \label{T.2.24}
\end{equation}
where $\Omega _n$ is a complex eigen value, and $|n+>$ are right
eigenvectors, and $<n-|$ left eigenvector of $h(out)$. Even if $h(out)$ is
hermitian, it has complex eigenvalues because we are using a new spectral
decomposition, which is only possible because we are working in a convenient
Gel'fand triplet. This fact will be the main tool that we will use below.
The eigenvalues and their squared will be written as: 
\begin{equation}
\Omega _n=\omega _n-\frac i2\gamma _n,.....\gamma _n\geq 0  \label{T.2.25}
\end{equation}
\begin{equation}
\Omega _n^2=m^2a(out)^2+z_n  \label{T.2.26}
\end{equation}

compared with eq. (\ref{T.2.12}) this equation shows that there are also
poles in the variable $k^2$ which are complex numbers.

\section{Time evolution.}

Coming back to the WKB expansion (\ref{WKB}), if we now consider the next
order and the time defined in eq. (\ref{T.2.7}), the function $\chi
(a,\varphi )$ must satisfy the Schroedinger equation: 
\begin{equation}
i\frac{d\chi }{d\eta }=h(\eta )\chi  \label{T.3.2}
\end{equation}
where $h(\eta )$ is hamiltonian $h$ written as a function of $\eta .$ Even
if this hamiltonian is time depending we can consider that for scales $%
a>a_{out}$ there is not particles creation and therefore we have an
invariant adiabatic vacuum $|0,out>$ and a definitive pole structure for the
S-matrix \cite{Lombardo}, so for $a(\eta )>a_{out}$ expansion (\ref{T.2.23})
will always have the same structure$.$ Thus the time evolution of $\chi $
will be: 
\begin{equation}
\chi (\eta )=\exp [-i\int h(\eta )d\eta ]\chi (0)  \label{T.3.3}
\end{equation}
From this equation we can obtain some conclusions:

i.- In particular the time evolution of the right eigenvector $|n+>$ reads: 
\begin{equation}
|n(\eta )+>=\exp [-i\int \Omega _n(\eta )d\eta ]|n(0)+>  \label{T.3.4}
\end{equation}
because even if the pole structure remains fixed the poles move as can be
seen in the example of ref. \cite{Lombardo}, eq. (3.3). So, from eq. (\ref
{T.2.25}) we can see that, if there are some $\gamma _n>0,$ some of these
eigenvectors have a dumped evolution$.$ Therefore, these eigenvectors
correspond to decaying states. Thus our formalism naturally yields decaying
states that vanish towards, the direction of time, that we can call the {\it %
future.}

ii.- Using eq. (\ref{T.Z.9}) we can expand any function $|\varphi >\in \phi
_{-}$as: 
\begin{equation}
|\varphi >=\sum_n|n+><n-|\varphi >+\int_0^\infty |\omega +><\omega -|\varphi
>d\omega  \label{T.3.5}
\end{equation}
then its time evolution will be: 
\[
|\varphi (\eta )>=\sum_n\exp [-i\int \Omega _n(\eta )d\eta ]|n+><n-|\varphi
>+ 
\]
\begin{equation}
+\int_0^\infty \exp [-i\int \Omega _k(\eta )d\eta ]|\omega +><\omega
-|\varphi >d\omega  \label{T.3.6}
\end{equation}
where all terms in the summatory, such that $\gamma _n\neq 0,$ have a
decaying evolution, while the rest of the terms and the integral have an
oscillatory behavior. So in the time evolution of (almost) any state we have
a decaying term that vanishes towards the future.

iii.- In this way the asymmetry introduced in hypothesis H$_2$ produces an
effective {\it time asymmetry}, because it allows to define a future time
direction. Moreover, it can be proved that, if in eq. (\ref{T.3.3}) the
evolution operator is considered as an operator from space $\phi_{-}$
to space $\phi _{-}$, namely if we restrict the dynamics to space $\phi
_{-},$ eq. (\ref{T.3.3}) is only defined for $\eta \geq 0.$ Therefore the
evolution operator cannot be inverted and so, it is really an {\it %
irreversible operator} (see \cite{Bohm}, \cite{CGG}, \cite{Laura}).

iv.- Let us consider the case of mixed states. For a mixed state $\rho \in
\Phi _{-}$ we can generalize the spectral decomposition (\ref{T.3.5}) to
obtain: 
\[
\rho =\sum_{n,m}\rho _{nm}|n+><n+|+\sum_n\int_0^\infty \rho _{n\omega
}|n+><\omega +|d\omega + 
\]
\begin{equation}
+\int_0^\infty \sum_n\rho _{\omega n}|\omega +><n+|d\omega +\int_0^\infty
\int_0^\infty \rho _{\omega \omega ^{\prime }}|\omega +><\omega +|d\omega
d\omega ^{\prime }  \label{T.3.6'}
\end{equation}
Repeating the computation of the pure state case, we can compute the time
evolution of state $\rho (\eta ).$ Since either $\gamma _n=0$ or $\gamma
_n>0 $ there will be oscillating terms and dumped ones then we obtain: 
\begin{equation}
\rho (\eta )=\rho _{*}(\eta )+\exp [-\frac 12\int \gamma d\eta ]\rho _1(\eta
)  \label{T.3.7}
\end{equation}
where the first term, of the r.h.s. is an oscillatory term and the second a
decaying term, where we have written a first factor corresponding to the
slowest dumping factor, namely $\gamma $ is the smallest of the non-zero $%
\gamma _n.$ When $\eta \rightarrow \infty $ we have: 
\begin{equation}
\rho (\eta )\rightarrow \rho _{*}(\eta )  \label{T.3.8}
\end{equation}

$\rho _{*}(\eta )$ is a thermodynamical equilibrium state. In fact, since in
its evolution there are not dumping factors, it behaves like an ordinary
stable quantum state and its entropy is time-constant, namely the one that
corresponds to thermodynamical equilibrium (below we will normalize this
constant to zero). It is logically a non stationary oscillatory equilibrium
state, because, even if it is in thermic equilibrium, the field cannot go to
dynamical equilibrium since, in our simple model, there are no interaction
terms among the field components. If these terms would be present, new
dumping factors would also be present and the final equilibrium would be a
stationary state \cite{E}.

This is the essence of our formalism. Below we will see the results that we
can obtain if we follow this road.

\section{Decoherence and correlations.}

In reference \cite{Lombardo}, using our formalism, it is proved that, if the
S-matrix has an infinite number of complex poles, we have decoherence and
that, in unstable states, configuration and momentum are correlated, in such
a way that the universe ends in a classical phase. In this demonstration
hypothesis H$_3$ plays an essential role. In paper \cite{Lombardo} it was
not proved that, in general, the S-matrix, relevant for our problem, has an
infinite set of complex poles, but that set was computed in one example,
while other examples were proposed.

Here we will not repeat the demonstration of ref. \cite{Lombardo}, but we
will complete this paper observing that using a potential, with a bounded
support, as in the present paper, the existence of an infinite set of poles
is a consequence of ref. \cite{Audrecht} and \cite{Nusenbein}.

In fact: A massive scalar field, conformally coupled, in metric (\ref{T.2.0}%
) satisfies Klein-Gordon equation: 
\begin{equation}
(\nabla _\mu \nabla ^\mu +m^2+\frac 16R)\psi =0,.....R=6a^{-3}\frac{\partial
^2a}{\partial \eta ^2}  \label{T.4.1}
\end{equation}
This equation leads, by variable separation to: 
\begin{equation}
\psi =\frac 1{(2\pi )^{\frac 32}a(\eta )}f(\eta )\exp (\pm ikx)
\label{T.4.2}
\end{equation}
and $f(\eta )$ satisfies a generalized oscillator equation with
time-depending frequency: 
\begin{equation}
f^{"}+\omega ^2(\eta )f=0,....\omega =[a^2(\eta )m^2+k^2]^{\frac 12}
\label{T.4.3}
\end{equation}
On the other hand, in ordinary quantum mechanics, the stationary solution $%
u(x)$ of a massive particle in a potential $W(x)$ satisfies: 
\begin{equation}
u"(x)+\omega ^2(x)u(x)=0,.....\omega (x)=[2m(E-W(x))]^{\frac 12}
\label{T.4.4}
\end{equation}
So both phenomena can be mathematically related according to the analogy: 
\begin{equation}
a^2(\eta )m^2\leftrightarrow -2mW(x),.....k^2\leftrightarrow 2mE
\label{T.4.5}
\end{equation}
More details about this analogy can be obtained from paper \cite{Audrecht}.

Now from reference \cite{Nusenbein}, page. 218, we know that the S-matrix of
a cut-off potential $W(x),$ namely a potential with a bounded support, has
an infinite number of complex poles. Our potential is $\sim $$a^2(\eta )$
which is practically a constant for $a\gg a_1,$ since $\varepsilon \cong 0,$
and exactly a constant if we consider the real value $\varepsilon =0$ so,
subtracting this constant final value, we can say that it is a cut-off
potential, with a S-matrix endowed with an infinite set of poles.

Almost all potentials used in the literature of Quantum Field theory in
Curved Space-time \cite{Birrell} are very well behaved in the infinities and
can be mimicked by this, bounded support, kind of potentials (this is not
the case for some QC potentials, that we will discuss in the conclusions).
So the existence of an infinite set of poles seems quite a general feature
of the theory. Thus, using the equation of paper \cite{Lombardo}, it can be
proved that our formalism leads to decoherence, to correlations, and to the
outcome of a classical universe, in a natural way.

Finally in particular subsystems of the universe the S-matrix has poles if
unstable quantum states exist in the subsystem \cite{Bohm}. Of course, these
poles will also appear in any complete S-matrix of the universe.

\section{Entropy}

Let $\rho (\eta )$ be the density matrix, of the universe or one of its
subsystems, for a physical admissible state ($\rho \in \Phi _{-})$ and let $%
\rho _{*}(\eta )$ be the corresponding thermodynamical equilibrium matrix.
In the universe these matrices are related by eq. (\ref{T.3.7}). The
hamiltonian of the subsystem is necessarily a term of the general
hamiltonian of eq. (\ref{T.2.2}) and its S-matrix must have poles if the
subsystem is not trivial (see papers \cite{CGG} and \cite{Bohm}). So we can
repeat all what we have said for the universe for the case of the subsystem
and we can also choose a t-asymmetric regular space state for the subsystem.
We must take care that the dumping or future direction of the subsystem
coincides with the dumping or future direction of the universe for
consistency. I. e. the local and global arrows of time must coincide. Then
we will also find eq. (\ref{T.3.7}) for the subsystem. The only difference
would be that, if the subsystem hamiltonian $h$ is no time dependent (with
respect e. g. to the proper time $t)$ in the integral of eq. (\ref{T.3.7})
must be subtituted by the usual product $ht.$

Then we can define the conditional entropy $S=$ $S[\rho (\eta )|\rho
_{*}(\eta )],$ of states $\rho (\eta )$ with respect to state $\rho
_{*}(\eta )$ \cite{Mackey} both for the universe or the subsystem: 
\begin{equation}
S=S[\rho (\eta )|\rho _{*}(\eta )]=-tr\{\rho (\eta )\log [\rho
_{*}^{-1}(\eta )\rho (\eta )]\}  \label{T.3.8'}
\end{equation}
such that $S[\rho _{*}(\eta )|\rho _{*}(\eta )]=0,$ namely the entropy
vanishes at equilibrium. We can, as well, use the corresponding classical
definition, since really we are interested in the classical phase of the
universe, but, in order to use only one notation, we will use the quantum
formulae.

From eq. (\ref{T.3.7}) we see that: 
\begin{equation}
tr\rho (\eta )=tr\rho _{*}(\eta )=1\Rightarrow tr\rho _1(\eta )=0
\label{T.3.9}
\end{equation}
i.e.: if the states $\rho (\eta )$ and $\rho _{*}(\eta )$ are normalized as
it should be, $\rho _1(\eta )$ has a vanishing trace, thus $\rho _1(\eta )$
is not a state but the coefficient of a correction of the equilibrium state
to obtain state $\rho (\eta ).$ The vanishing of the trace of $\rho _1$ is
directly proved in papers \cite{CGG},\cite{Laura}, \cite{Gadella}, using our
formalism. Now if we expand the logarithm in eq. (\ref{T.3.8'}), and use
eqs. (\ref{T.3.9}) we obtain: 
\begin{equation}
S=S[\rho (\eta )|\rho _{*}(\eta )]=-\exp [-\frac 12\int \gamma d\eta
]tr[\rho _{*}^{-1}(\eta )\rho _1^2(\eta )]+...  \label{T.3.10}
\end{equation}
where the dots symbolize higher order terms \cite{Lambda}.

This entropy has the property: 
\begin{equation}
\lim_{\eta \rightarrow \infty }S[\rho (\eta )|\rho _{*}(\eta )]=0
\label{T.3.11}
\end{equation}
namely the entropy evolves towards its null equilibrium value. This is so
because the prefactor in eq. (\ref{T.3.10}) dominates any other time
variation, since $\rho _{*}(\eta )$ is usually oscillatory (namely it will
be oscillatory in the case a, but not in the case b, see below) and $\rho
_1(\eta )$ has oscillatory terms and dumping factors that vanish faster than
the dominant decaying factor.

In eq. (\ref{T.3.8'}) we have two matrices: $\rho (\eta )$ and $\rho
_{*}(\eta )$, then we have also two possibilities. Either both matrixes have
the same kind of evolution or they have different ones. The first is the
general case, but the second case appears when $\rho _{*}(\eta )$ follows a
different evolution due to, e.g., an external agency. Let us consider the
two cases:

a.- Both matrices follow the same evolution law. Namely, if we have a time
variable hamiltonian $h(\eta )$, as in the case of the universe, the
evolution will be: 
\begin{equation}
\rho (\eta )=\exp [-i\int_{\eta ^{\prime }}^\eta l(\eta )d\eta ]\rho (\eta
^{\prime }),........\rho _{*}(\eta )=\exp [-i\int_{\eta ^{\prime }}^\eta
l(\eta )d\eta ]\rho _{*}(\eta ^{\prime })  \label{T.3.12'}
\end{equation}
where $l(\eta )$ is the corresponding Liouville operator, i. e., $l(\eta
)\rho =h(\eta )\rho -\rho h(\eta )$.

In subsystems of the universe where the hamiltonian $h$ is not a proper-time
variable( e. g. in subsystem which does not expand or contract due to an
external agency), we would have: 
\begin{equation}
\rho (t)=\exp [-il(t-t^{\prime })]\rho (t^{\prime })...........\rho
_{*}(t)=\exp [-il(t-t^{\prime }))]\rho _{*}(t^{\prime })  \label{T.3.12}
\end{equation}
where $l$ is the Liouville operator corresponding to hamiltonian $h$ and $t$
is the proper time$.$ In this case the equilibrium matrix $\rho _{*}(t)$
also evolves in the same way than matrix $\rho (t).$ Now, from what we have
said in Sec.3, point iii, eq. (\ref{T.3.12'}) is only valid if $\eta >\eta
^{\prime }.$ Analogously, if the subsystem S-matrix has poles and we have
also chosen for the subsystem a similar admissible function space $\Phi
_{-}, $ eq.(\ref{T.3.12}) is only valid if $t-t^{\prime }\geq 0$ or $t\geq
t^{\prime }.$ Therefore the last two evolutions are irreversible. Now, from
ref. \cite{Mackey} (using the classical-quantum analogy since we are in the
classical phase, the ''$\exp "$ operator of the last two equations will be a
Frobenius-Perron operator, and we can use the classical definition of
conditional entropy) we know that: 
\begin{equation}
S[\rho (\eta )|\rho _{*}(\eta )]\geq S[\rho (\eta ^{\prime })|\rho _{*}(\eta
^{\prime })],...S[\rho (t)|\rho _{*}(t)]\geq S[\rho (t^{\prime })|\rho
_{*}(t^{\prime })]  \label{T.3.13}
\end{equation}
respectively. It would be = if the evolution operators would be reversible, 
{\it but they are not }(consider also eq.( \ref{T.3.10})). Then these
entropies are really monotonically growing. Therefore we have proved the
second law of thermodynamics, for the whole universe or for any non trivial
subsystem. So our formalism yields this fundamental law naturally (compare
with the much more complicated coarse graining method of paper \cite
{Scochimarro}).

The demonstration is based in the fact that (in both cases) $\rho $ is an
admissible state (like the ink drop spreading in the glass of water), so $%
\rho \in \Phi _{-}$ and $\gamma _n\geq 0.$ If we would have taken $\rho \in
\Phi _{+}$ it would be $\gamma _n\leq 0$ and the entropy would decrease
showing that:

i.- In the case of the subsystem. $\Phi _{+}$ is the space of non admissible
solution (the ink drop contracting spontaneously). In fact, in this case the
arrow of time is the one of the universe and not the one of the subsystem,
and in the subsystem we will see a decaying of the entropy, showing that the
state is not physically admissible.

ii.- In the case of the universe. Going from $\Phi _{-}$ to $\Phi _{+}$ we
have simply changed our convention (since all possible arrows of time are
embodied in the universe), calling the ''future'' the direction of
decreasing entropy.

b.- Let us now consider the important case of a subsystem of the universe:
the matter and radiation within an expanding universe ( i. e. we do not take
into account the entropy of the gravitational field, and we consider that
this filed as an external agency that expands the space, where the matter
and the radiation are located). Then the conditional entropy is not
necessarily monotonically increasing, at least for short times. In fact, we
cannot use eq. (\ref{T.3.13}) since $\rho _{*}(\eta )$ does not satisfy an
equation like (\ref{T.3.12'}$_2),$ because its evolution is fixed by the
external universe expansion. This is completely logical since $S[\rho (\eta
)|\rho _{*}(\eta )]$ is just the matter radiation entropy (with no
gravitational field entropy contribution) in an expanding (or contracting)
universe with a equilibrium state $\rho _{*}(\eta ),$ that varies in an
independent way and it is well known that matter (let us say a gas) can have
decreasing entropy into a variable geometry (let us say a box with moving
walls). A phenomenological study of the problem can be found in ref. \cite
{Davies}, \cite{Aquilano}. In this case what we have called up to now $S$ is
just the entropy gap $\Delta S$ with respect to a variable maximal possible
entropy $S_{\max }.$ The actual entropy, which grows monotonically, is $%
S=S_{act}=S_{\max }+\Delta S$ . But $\Delta S$ has not this property.
Furthermore, the diminishing of $\Delta S$ for short times is welcome, as we
will see in the next two sections.

Finally let us consider the origin of the ''miracle'' that allows us to
define a growing entropy with no coarse-graining. The miracle is produced by
the generalized spectral decompositions (\ref{T.3.5}) or (\ref{T.3.6'})
which yield the evolution (\ref{T.3.6}) with only dumping (and oscillatory)
factors, which, in turn, are only possible if we introduce an asymmetry,
like the one of hypothesis H$_2.$ Namely if we restrict the regular state
space in a t-asymmetric way we obtain also the asymmetric generalized states
of expansions and eqs. (\ref{T.3.5}) and (\ref{T.3.6'}). Is this fact so
strange? Experimentally we know that we only have a finite set of physical
measurements to fix a quantum state, e. g. we know the value of the wave
function in a finite set of point of the configuration space $Q$. In the
limit we will have a set of discrete data that allows us to define by
interpolation e. g. a polynomial belonging to the set of polynomials ${\cal P%
}$ of the introduction. But in order to work we need the whole wave function
complete superspace and therefore we must complete the space expanded by the 
${\cal P}$. In the ordinary theory we complete the space expanded by the $%
{\cal P}$ with the topology of the norm of ${\cal H}$ or the nuclear one of $%
{\cal S.}$ Up to this point we have a time-symmetric theory. If we want to
introduce time asymmetry we coarse-grain the system, to obtain a new
complete space of relevant states. In the new theory we complete using
Hardy-Schwartz function (i. e. we complete with the nuclear topology of $%
\phi _{-})$. So we only use one step, instead of two, and obtain the same
physical results. But, really both methods have essentially the same
physical base: we have only a finite (or discrete) amount of information and
this information must be worked out, somehow, to obtain the complete space
of regular states of the theory. So, as expected, there is no miracle.

\section{The entropy gap.}

In this section we study the universe entropy gap $\Delta S=S_{act}-S_{\max
} $, following a qualitative idea of Paul Davies \cite{Davies}. We will
complete this idea actually computing the entropy gap after decoupling time.
Therefore we change our model, it will be still homogeneous and isotropic,
with metric ( \ref{T.2.0}), but obviously the particle production is
finished, so we will consider that we are simply in a flat geometry, matter
dominated, universe.

It is well known that the isotropic and homogeneous expansion of the
universe is a reversible process with constant entropy \cite{Tolman}. In
this case the matter and the radiation of the universe are in a thermic
equilibrium state $\rho _{*}(t)$ at any time $t$. As the radiation is the
only important component, from the thermodynamical point of view, we can
choose $\rho _{*}(t)$ as a black-body radiation state \cite{COBE}, i. e. $%
\rho _{*}(t)$ will be a diagonal matrix with main diagonal: 
\begin{equation}
\rho _{*}(\omega )=ZT^{-3}\frac 1{e^{\frac \omega T}-1}  \label{1}
\end{equation}
where $T$ is the temperature, $\omega $ the energy, and $Z$ a normalization
constant (\cite{Landau}, eqs. (60.4) and (60.10)). The total entropy is: 
\begin{equation}
S=\frac{16}3\sigma VT^3  \label{2}
\end{equation}
(\cite{Landau}, eq. (60.13)) where $\sigma $ is the Stefan-Boltzmann
constant and $V$ a commoving volume.

Let us consider our isotropic and homogeneous model of universe with scale $%
a.$ Any commoving volume evolves as $V\sim a^3,$ and, since from the
conservation of the energy-momentum tensor and radiation state equation, we
know that $T\sim a^{-1},$ we can verify that $S=const.$ Thus the
irreversible nature of the universe evolution is not produced by the
universe expansion, even if $\rho _{*}(t)$ has a slow time variation.

Therefore, after decoupling time, the main process that has an irreversible
nature is the burning of unstable $H$ in the stars (which produces $He$ and,
after a chain of nuclear reactions, $Fe$). This unstable state produces
poles in the corresponding S-matrix and a nuclear reaction process, with
mean life-time $t_{NR}=\gamma ^{-1}$ . Therefore, using eq. (\ref{T.3.7}),
and considering that $\gamma $ is constant (under proper time variation),
since it corresponds to a local process considered in the paragraph ''a'' of
the last section, (or simply on phenomenological grounds) we can then say
that the state of the universe, at time $t$, is: 
\begin{equation}
\rho (t)=\rho _{*}(t)+\rho _1e^{-\gamma t}+0[e^{-\gamma t}]  \label{3}
\end{equation}
where $\rho _1$ is a certain phenomenological coefficient, which is constant
in time since all the time variation of nuclear reactions is embodied in the
exponential law $e^{-\gamma t}$. Also on phenomenological grounds, we can
foresee that $\rho _1$ must peak strongly around $\omega _1$, the
characteristic energy of the nuclear process. All these reasonable
phenomenological facts can also be theoretically explained in different
ways, e. g.: Eq. (\ref{3}) can be computed with the theory of paper \cite
{Sudarshan} . In reference \cite{Laura} it is explicitly proved that $\rho
_1 $ peaks strongly at the energy $\omega _1$. So using eq.(\ref{T.3.8'}) we
can compute the entropy gap:

\begin{equation}
\Delta S=-tr[\rho \log (\rho _{*}^{-1}\rho )]  \label{5}
\end{equation}
Using now eq. (\ref{3}), and considering only times $t\gg t_{NR}=\gamma
^{-1} $ we can expand the logarithm, as in eq. (\ref{T.3.10}), to obtain: 
\begin{equation}
\Delta S\approx -e^{-\gamma t}tr\left( \rho _{*}^{-1}\rho _1^2\right)
\label{6}
\end{equation}
where we have used eq. (\ref{T.3.9}). We now introduce the equilibrium state
(\ref{1}) for $\omega \gg T$ . Then: 
\begin{equation}
\Delta S\approx -Z^{-1}T^3e^{-\gamma t}tr(e^{\frac \omega T}\rho _1^2)
\label{7}
\end{equation}
where $e^{\frac \omega T}$ is a diagonal matrix with this function as
diagonal. But as $\rho _1$ is peaked around $\omega _1$ we arrive to a final
formula for the entropy gap: 
\begin{equation}
\Delta S\approx -CT^3e^{-\gamma t}e^{\frac{\omega _1}T}  \label{8}
\end{equation}
where $C$ is a positive constant.

Let us now compute the time evolution of the entropy gap. We have computed $%
\Delta S$ for times larger than decoupling time and therefore, as $a\sim
t^{\frac 23}$ and $T\sim a^{-1},$ we have:

\begin{equation}
T=T_0\left( \frac{t_0}t\right) ^{\frac 23}  \label{9}
\end{equation}
where $t_0$ is the age of the universe and $T_0$ the present temperature.
Then: 
\begin{equation}
\Delta S\approx -C_1e^{-\gamma t}t^{-2}e^{\frac{\omega _1}{T_0}\left( \frac{%
t_0}t\right) ^{\frac 23}}  \label{10}
\end{equation}
where $C_1$ is a positive constant. Drawing the corresponding curve \cite
{Aquilano} it can be seen that $\Delta S$ has a maximum at $t=t_{cr_1}$ and
a minimum at $t=t_{cr_2}.$ Let us compute these critical times. The time
derivative of the entropy reads: 
\begin{equation}
\stackrel{\bullet }{\Delta S}\approx \left[ -\gamma -2t^{-1}+\frac 23\frac{%
\omega _1}{t_0T_0}\left( \frac{t_0}t\right) ^{\frac 13}\right] \Delta S
\label{11}
\end{equation}
This equation shows two antagonic effects. The universe expansion effect is
embodied in the second and third terms in the square brackets, being an
external agency to the matter-radiation system such that, if we neglect the
second term, it tries to increase the entropy gap and, therefore, to take
the system away from equilibrium (as we will see the second term is
practically negligible). On the other hand, the nuclear reactions embodied
in the $\gamma $-term, try to convey the matter-radiation system towards
equilibrium. These effects become equal at the critical times $t_{cr}$ such
that: 
\begin{equation}
\gamma t_0+2\frac{t_0}{t_{cr}}=\frac 23\frac{\omega _1}{T_0}\left( \frac{t_0%
}{t_{cr}}\right) ^{\frac 13}  \label{12}
\end{equation}
For almost any reasonable numerical values this equation has two positive
roots: $t_{cr_1}\ll t_0\ll t_{cr_2}$. Precisely:

i.- For the first root we can neglect the $\gamma t_0$-term and we obtain: 
\begin{equation}
t_{cr_1}\approx t_0\left( 3\frac{To}{\omega _1}\right) ^{\frac 32}
\label{13}
\end{equation}
(this quantity, with minus sign, gives the third unphysical root).

ii.- For the second root we can neglect the $2(t_0/t_{cr})-$term, and we
find: 
\begin{equation}
t_{cr_2}\approx t_0\left( \frac 23\frac{\omega _1}{T_0}\frac{t_{NR}}{t_0}%
\right) ^3  \label{14}
\end{equation}

Let us make now some numerical estimates. We must choose numerical values
for four parameters: $\omega _1=T_{NR},$ $t_{NR}=\gamma ^{-1},$ $t_0,$ and $%
T_0.$

$T_{NR}$ and $t_{NR}$ can be chosen between the following values \cite{Cumul}%
: 
\begin{equation}
T_{NR}=10^6..to..10^8{}.^0K  \label{15}
\end{equation}
\[
t_{NR}=10^6..to..10^9.years 
\]
while for $t_0$ and $T_0$ we can take: 
\begin{equation}
t_0=1.5\times 10^{10}.years  \label{16}
\end{equation}
\[
T_0=3^0K 
\]
In order to obtain a reasonable result we choose the lower bounds for $T_{NR}$ and $t_{NR}$ and for $t_{cr_1}$ we obtain $:$
\begin{equation}
t_{cr_1}\approx 1.5\times 10^3.years  \label{17}
\end{equation}
So $t_{cr_1}$ is smaller than the decoupling time and it should not be
considered since the physical processes before this time are different than
those we have used in our model. Also, we must only consider times $%
t>t_{NR}=\gamma ^{-1},$ in order to use eq. (\ref{6}).

For $t_{cr_2}$ we obtain: 
\begin{equation}
t_{cr_2}\preceq 10^4t_0  \label{18}
\end{equation}
From eqs. (\ref{17}) and (\ref{18}) we can see that really $t_{cr_1}\ll
t_0\ll t_{cr_2}.$ Thus:

-From $t_{NR}$ to $t_{cr_2}$ the expansion of the universe produces a
decreasing of entropy gap, according to Paul Davies prediction \cite{Davies}%
. Also, it probably produces a growing order, and therefore the creation of
structures like clusters, galaxies and stars \cite{Reeves}.

-After $t_{cr_2}$ we have a growing of entropy, a decreasing order and a
spreading of the structures: stars energy is spread in the universe,which
ends in a thermic equilibrium \cite{AJP}. In fact, when $t\rightarrow \infty 
$ the entropy gap vanishes (se eq. (\ref{10})) and the universe reaches a
thermic equilibrium final state.

$t_{cr_2}\preceq 10^4t_0$ is the frontier between the two periods. Is the
order of magnitude of $t_{cr_2}$ a realistic one? In fact it is, since $%
10^4t_0\approx 1.5\times 10^{14}years$ after the big-bang all the stars will
exhaust their fuel \cite{AJP} , so the border between the two periods should
have this order of magnitude. Further, it should also be smaller than this
number. This is precisely the result of our calculations (see also letter 
\cite{Aquilano}).

So we are at the edge of a correct physical prediction, even if our model is
extremely naive and simplified: an homogeneous universe and besides we have
neglected the higher order terms in eq. (\ref{3}) which perhaps may be
important for finite times. Besides in the real universe nuclear reactions
take place within the stars, that can only be properly considered in a
inhomogeneous geometry. Nevertheless, this rough numerical estimate shows
the consistency of all the theory. Furthermore the decreasing of the entropy
gap, in the period $t_{NR}<t<t_{cr_2}$, will be crucial in the next section.

\section{The branch system.}

The set of {\it irreversible} processes within the universe, each one
beginning in an unstable non-equilibrium, state can be considered a {\it %
branch system }\cite{Davies}{\it , }\cite{Reichenbach}{\it , . }Namely,
every one of these processes begin in a non-equilibrium state, such that,
this state was produced by a previous process of the set. E. g.: Gibbs ink
drop (initial unstable state) spreading in a glass of water ( irreversible
process) is only probable (since the probability to create an ink drop by
fluctuations is extremely small) if there was first an ink factory, which
extracted the necessary energy from an oven, where coal (initial unstable
state) was burnt (branched irreversible process); in turn coal was created
with energy coming from the sun, where $H$ (initial unstable state) is burnt
into $He$ (branched irreversible process); finally $H$ was created using
energy obtained from the unstable initial state of the universe (the
absolute initial state of the branch system). Therefore, using this
hierarchical chain, all the irreversible processes are related to the
cosmological initial condition, the only one that must be explained. Let us
observe that:

i.- The branch system defines its own arrow of time the {\it branch arrow of
time (BAT),} as the direction that goes from the unstable initial state of
every member of the system towards equilibrium. Probably the BAT is the most
useful of all arrows of time, since it is present in any irreversible local
process.

ii.- Once we have the branch system the irreversible evolution of each
system is easy to explaining, since once we have understood the origin of
the initial unstable state of each irreversible process within the universe
(even if we have not yet discussed the origin of the initial state of the
whole universe) it is not difficult to obtain Lyapunov variables (or
irreversible evolution equations), if we consider, e. g. that the subsystems
where these processes take place are not isolated. If it is so, forces of
stochastic nature penetrate from the exterior of each subsystem and, it is
well known, that if we add stochastic terms to time-symmetric evolution
equation, we obtain time-asymmetric ones, yielding Lyapunov variables e. g.
a non-decreasing entropy \cite{Mackey}. We can also consider that each
subsystem has an enormous amount of information and we are able to measure,
compute, and control a part of this information, that we will call {\it %
relevant.} If we neglect the rest of the information, the {\it irrelevant }%
one, we can obtain also irreversible evolution equations and Lyapunov
variables \cite{Mackey},\cite{Irre}. These two procedures can be considered
within the coarse-graining usual formalism.

iii.-But of course, if we follow the ideas of this paper, we will use more
refined mathematical tools, and in each of the subsystem, introduce a model
similar to the one we have used for the whole universe, as we have done in
section 5, introducing the hypothesis H$_2$ in each subsystem. (It has
already been done in papers \cite{CGG}, \cite{Bohm}, \cite{Laura}, and the
same results are obtained, i. e. irreversible evolution equation, Lyapunov
variables, etc.) Then we see that entropy grows in each subsystem provided
the state of the subsystem would be chosen among the physically admissible
states of space $\Phi _{-}.$ Then each subsystem of the branch system begins
in an unstable, low entropy, state and evolves towards thermal equilibrium.
The physical non-admissible states (those of space $\Phi _{+}$) correspond
to theoretical evolutions that would only exist before the instant of
creation of the subsystem (the instant when we put the ink in the glass of
water ) evolving with decreasing entropy, towards that instant (namely the
ink drop contracting spontaneously) . These evolutions simply do not exist
in nature because, before the instant of its creation, the subsystem really
does not exist as such. Before that instant a different subsystem exits with
different evolution laws (the ink factory that creates the ink drop).
Therefore all the scenario turns out to be realistic and satisfactory.

iv.- So only one problem is left: Why did the universe begin in an unstable
low-entropy state? Let us first observe that really we are referring not to
the ''whole'' universe (with its gravitational field) but only to the
matter-radiation subsystem of the universe.

In the no-time version of the introduction we have postulated H$_1$ and H$%
_2. $ Using these hypotheses we have reconstructed time and demonstrated, in
the sections above, that the universe expansion creates, in its
matter-radiation subsystem, an entropy gap $\Delta S$ that takes it out of
equilibrium, not only at $t=0$, but in a long period of its history, since
the actual entropy is $S_{act}<S_{\max }$. We have also demonstrated that
the matter-radiation subsystem of the universe evolves to a final state of
thermic equilibrium since $\Delta S\rightarrow 0,$ when $t\rightarrow \infty 
$ (cf. eq. ( \ref{T.3.11})). So the answer to the only problem left is
essentially hypothesis H$_{2.}$ Certainly, someone will think that we have
solved a problem by postulating an axiom, and this is not a very exiting
result. But if the axiom yields the solutions of many problems, and this is
the case of H$_2,$ the axiom must be welcome. After all, this is the role of
axioms.

v.- Finally we can ask ourselves if, in the perspective of the branch system
idea, H$_2$ is a natural hypothesis. H$_2$ says that $\rho \in \Phi _{-}\in 
{\cal S\otimes S}$. So first it is postulated ,that $\rho \in {\cal S\otimes
S,}$ and therefore $\rho $ is a smooth function, with infinite derivatives,
and well behaved in the infinities of the configuration space. This part of H%
$_2$ seems quite natural. Certainly much more natural than the two other
alternative possibilities:

a.- $\rho \in {\cal L,}$ in which case $\rho $ can be, e. g., a square
integrable function of ${\cal H\otimes H,}$ where in a set of points, the
function can take non continuous an arbitrary values. What is the physical
meaning of this discontinuity?

b.- $\rho \in {\cal S^{\times }\otimes S^{\times },}$ namely a distribution,
e. g. a delta function, certainly a quite unnatural state.

So the first part of H$_2$ is natural. The second part is to ask why $\rho $
would be endowed with a natural asymmetry, the one of $\Phi _{-}.$ Is it too
much to ask? Let us study this question according to the branch system idea
and our formalism. There will be no branch system only if the universe (and
now we are referring to the whole universe with the gravitational field
included) would begin in an equilibrium state $\rho _{*}$, since in this
case it will always remain in equilibrium. Now, from eqs.(\ref{T.3.6'}), (%
\ref{T.3.7}), and (\ref{T.3.8}) we see that, in this case $\rho _{*}\in \Phi
_{-}^{\times },$ so $\rho _{*}$ would be a distribution \cite{Baker},
something like a delta function, and we have just considered this choice as
unnatural. On the contrary $\rho \in \Phi _{-}$ is a much more natural
state. Any state of $\Phi _{-}$ will produce a branch system, since any
state of $\Phi _{-}$ yields eq. (\ref{T.3.7}). So we can, at least, conclude
that H$_2$ is the requirement that the state of the universe id a natural
and an asymmetric one. H$_2$ is also intimately related with the branch
system idea and in consequence it is also related with the fact that really
our universe is a branch system. H$_2$ is just the transcription of these
physical facts.

\section{Coordination of the arrows of time.}

In this section we will only consider the coordination of the arrows of time
related with our model, namely:

\begin{itemize}
\item  {\it The branch arrow of time (BAT), }the arrow that goes from the
unstable initial state of every process of the universe branch system to its
equilibrium final state. As we have seen in the last section this arrow is a
direct consequence of the asymmetry introduced by H$_{2.}$

\item  {\it The thermodynamic arrow of time (TAT), }that points to the
direction of the growing of the universe entropy $S.$

\item  {\it The cosmological arrow of time (CAT), }that points to the
direction of the growing of the universe scale $a.$
\end{itemize}

Of course, all these arrows are related with time and therefore they must
only be considered in the classical (or semiclassical) period where the time 
$\eta $, given by eq. (\ref{T.2.7}), is well defined. In the timeless
quantum period we only have the asymmetry defined by H$_2.$

Then: eq. (\ref{T.3.13}), for non-expanding or contacting subsystems within
the universe, which is a consequence of H$_2$, shows that BAT=TAT, and that $%
t$, or more generally $\eta $, grows in the same direction than $S$.

The relation between BAT(=TAT) and CAT is given by the $\pm $ sign in eqs. (%
\ref{T.2.8}) or (\ref{T.2.9}). Then these two arrows of time, a priori, are
not coordinated in our model. But in the classical period we have just {\it %
one} classical universe and therefore the $\pm $ sign and the constant $C$
are fixed$,$ so in the classical period we have just one sign: either + or -
(see App. 12). Therefore, once the sign is fixed, a clear relation appears
between BAT and CAT:

\begin{itemize}
\item  If the model is an expanding one (and we choose the sign +) we will
have BAT=CAT, at least in the final evolution where eq. (\ref{T.2.9}) is
valid (if we make the unusual choice of the - sign we are just changing the
conventional direction of time $\eta ,$ with no physical consequences).

\item  If the model is an expanding-contracting one (and we choose the sing
+) we will have BAT=CAT in the expanding period and BAT$\neq $CAT in the
contracting period. BAT=CAT is, in fact, the definition of the expanding
period and BAT$\neq $CAT is the definition of the contracting one. But
BAT=TAT does not change when we go from the expanding to the contracting
period or vice versa, since the choice of $\Phi _{-}$ (or $\Phi _{+})$ is
made once and for all.
\end{itemize}

So the study of the correlation of the arrows of time is completed, and
almost trivial, because we have H$_2$ that defines BAT. (See also \cite{QAT})

\section{Other results.}

The main results related with quantum cosmology are stated in the above
sections. But we must comment that using the present formalism all the
relevant results of irreversible statistical mechanics can also be obtained,
e. g. all the results of the book \cite{Balescu}, as it is proved in ref. 
\cite{CGG}, because the main $\Pi $ projector of the quoted book can be
defined using Gel'fand triplets. Also, in some simple cases, we can go from
the quantum models to the classical ones \cite{Diener}, where we find the
same philosophy, in classical cases. Chaotic models like Baker's
transformation and Renyi's maps, are also treated with the same method, with
good results \cite{Antonioutasaki}. Other interesting results are contained
in papers \cite{Chiu}, \cite{Bohm}, \cite{Sudarshan},\cite{Antoniou}, , and 
\cite{QAT}. So what we have explained is just the quantum cosmological
chapter of a general method to deal with irreversible processes.

\section{Conclusions.}

Let us summarize our main conclusions.

i.- All our scheme is based in the existence of a physically admissible
state superspace $\Phi _{-}$ and of a physically forbidden state superspace $%
\Phi _{+}.$ Thus, the time inversion that goes from $\Phi _{-}$ to $\Phi
_{+} $ is also forbidden. Namely, no Maxwell demon can change the direction
of all the velocities of the universe. This is, of course a practically
impossible task. Is it also theoretically impossible? In fact it is, even if
Maxwell demon would change all these directions (while we are sleeping) we
will not notice the change (when we wake up), since {\it all arrows of time
would be changed} and we would not have any extra arrow to verify the
change. Thus this global-demon task is conceptually impossible.

ii.- What we have presented is not a mathematical theorem, but a model that
can be generalized in many ways. These more general models will have a
similar behavior, than the present one, if two essential features are
present: the existence, at the quantum gravity level, of a geometry of
maximal relative probability which allows us to construct ''out'' states for
the fields, and a S-matrix with infinite complex poles. The first
requirement seems natural for any realistic model of universe. On the other
hand we have restricted the class of possible potentials in order to be able
to prove that the corresponding S-matrix has infinite complex poles. But
several QC potentials do not belong to this class, because they have a bad
behavior at infinity. Nevertheless, usually, they also have an infinite set
of poles, as it can be proved case by case \cite{Diener}. So the two basic
features seem usual enough to consider that our model is a good sample of
the general behavior of the universe. Then we can say that:

iii.- If we introduce an adequate regular state space (or an adequate
topology) it seems that all the known results of statistical irreversible
physics can be reobtained. It must be emphasized that we are {\it not adding
a new object to the theory, since a regular state space (or the
corresponding topology) must be defined anyhow. }We are just choosing the
most convenient one. Let us repeat the general relativity example: The
space-time has a metric, we can choose a flat space-time metric or a curved
one. In the second case we explain gravity in the best possible way. We add
nothing, we just choose the best mathematical structure. The same thing
happens in the present case. If we choose the usual regular state space $%
{\cal S}$ we are forced to make a coarse graining (and there is nothing
experimentally wrong with coarse graining, as there in nothing
experimentally wrong with post-newtonian theories, but both are ''non
economical'' formalisms). If we choose the new regular state space $\phi
_{-} $ we make two steps in one, so we have a conceptual advantage.

iv.- Precisely, because the new formalism is conceptually clearer we can see
that time-asymmetry is just a kind of spontaneous symmetry breaking.

v.- Most probably the old and new formalism will always yield the same
physical results, because they are both based in the same physical base: the
limited amount of information must be somehow worked out to obtain a
complete theory. Therefore, most likely, they are as experimentally
equivalent as general relativity and post-newtonian gravity with an infinite
number of terms.

So, even if we have not found any new or spectacular result, we think that
the introduced formalism presents a quite coherent picture of the real
time-asymmetric universe and shows us how time-asymmetry forces us to choose
a Gel'fand triplet as the mathematical structure of the theory.

\section{Appendix. Time-asymmetry of the regular states superspace.}

It is a common lore among physicists that, since the fundamental evolution
equations are time-symmetric and the universe is time asymmetric, therefore
the state of the universe must be the responsible for time asymmetry. Namely
time-asymmetry is created as in the case of spontaneous symmetry breaking.
The problem is to mathematically precise this idea. That is what we have
done in this paper, by showing that the asymmetry can be hidden in the
choice of an asymmetric regular state superspace. In fact, beginning with
the one symmetric regular state space ${\cal S}$ of the primitive theory, we
have introduced the asymmetry related with the lower and upper halfplanes,
of the complex plane, and we have defined asymmetric regular state spaces $%
\phi _{-}$ and $\phi _{+}$, and we have chosen one of them, say $\phi _{-},$
creating a symmetry breaking.

But, since this is perhaps the most important point, we will try to explain
this fact in the most didactical way. First let us consider that $a\geq 0,$
and let us draw the potential $V(\eta )=V[a(\eta )]$ in the right side of
the horizontal $\eta $ axis. Let us take $C$ such that $\eta =0$ would
correspond to $a=0$ in eq. (\ref{T.2.9}), to make the origin of $a$ to
coincide with the one of $\eta $. As, in principle, $\eta $ can be positive
or negative we will complete the previous figure with a symmetrical curve in
the left side of the vertical axis. Really as we cannot go from negative $%
\eta $ to positive ones we can add an infinite potential barrier at $\eta
=0. $ Now we have a more or less typical potential figure for a spontaneous
symmetry breaking. In our case the minima are unimportant (even if, most
likely, the equilibrium states will correspond to the minima in almost every
model) but since we have an infinite potential barrier at $\eta =0$ it is
clear that the universe can either be in the right or the left side of the
vertical axis. Choose either one or the other possibility. Then, you have
chosen the + or the - sign in eqs. (\ref{T.2.8}) or (\ref{T.2.9}) (since we
have chosen $a=0$ when $\eta =0$ and it is always $a\geq 0$). Let us
consider, for a moment the expanding case: if you choose + then BAT=CAT and
you have chosen the convention of all the formulae of this paper, namely you
have chosen a H$_2$ with a $\Phi _{-}$ space as in this paper. If you choose
the - sign, you just change the convention used in this paper, instead of $%
\eta $ you will have $-\eta $ everywhere and in H$_2$ you will have $\Phi
_{+}$. Of course, physically nothing is changed since you have just changed
a convention. In the expanding-contracting case we can repeat the reasoning
for the expanding phase only and everything turns out to be the same.

So we are clearly in a case of a symmetry breaking, produced by the state
space and not by the equations, where we have conventionally chosen between
two conventional different possibilities, either the right side of the
figure or the left side, namely between $\Phi _{-}$ and $\Phi _{+}.$ After
this choice the time-asymmetry appears but, of course, the choice itself is
irrelevant, as in every spontaneous symmetry breaking. So really there are
two $\pm $ choices in the theory: the one that corresponds to the choice
between $\Phi _{-}$ and $\Phi _{+}$ and the one in eqs. (\ref{T.2.8}) and (%
\ref{T.2.9}). We will explain more about the second choice in the next
appendix.

\section{Appendix. The asymmetry of Q space.}

Why do we have + and not - in eqs. (\ref{T.2.8}) or (\ref{T.2.9}) ? Perhaps
it is interesting to relate the present formalisms with the final
speculations of the second paper \cite{Barbour}:

If $\Phi _{-}$ is chosen (namely the first $\pm $ choice is made)and $C>0:$
Are the + and -, of eq. (\ref{T.2.9}) conventionally identical? Is the
choice between them just conventional or is there something substantially
different between them? Of course the difference between + and - is just
impossible to find in the quantum gravity timeless period. But the arena of
time asymmetry is the classical period where the $\pm $ sign and the
constant $C$ are fixed (and as in this case we have taken $C>0,$ as we have
said). Can we find a difference between + and - in this period? Our strategy
is to maintain the symmetry the WDW equation and create the time-asymmetry
using the states. In doing so we can use all the features of the states and
the configuration space $Q.$ And it turns out that $Q$ is not so symmetric
since it has boundaries \cite{Barbour} that gives $Q$ some peculiar
structure. Precisely $a\geq 0$ in our very simple model. So once the choice
of the space of state is made, using all the well known philosophy of
symmetry breaking, as explained in section Appendix 11, + and - become
obviously different in expanding models since:

\begin{itemize}
\item  With the + we would have equilibrium towards $a\rightarrow \infty $,
which is possible.

\item  With the - we would have equilibrium towards $a\rightarrow -\infty ,$
which is impossible since $a\geq 0.$
\end{itemize}

So we must choose the +. If we would have begun choosing $\Phi _{+}$ (and $%
C<0)$ we will be forced to choose the - sign. So the choice of the regular
state space implies the choice of a sign. In expanding-contacting models
things are more difficult since we do not have either $a\rightarrow \infty $
or $a\rightarrow -\infty $, so theoretically we never reach the equilibrium
state. In these models we must change the notion of equilibrium, and
consider that this state is obtained in long but finite times. Then, we must
consider only the expanding phase and repeat the reasoning of expanding
models.

Therefore, as it was already anticipated in ref. \cite{Barbour} the boundary
structure or $Q$ creates a difference, and allows us to choose the right
sign.

\section{Acknowledgments.}

I would like to warmly thank R. Aquilano, J. Barbour, and M. Gadella for
many stimulating discussions and to several colleagues, for a great number
of interesting questions, that I try to answer in the references.

This work was partially supported by grants: CI1$^{*}$-CT94-0004 of the
European Community, PID-0150 of CONICET (National Research Council of
Argentina), EX-198 of the Buenos Aires University, and 12217/1 of
Fundaci\'{o}n Antorchas and the British Council.


\begin{thebibliography}{99}
\bibitem{Lumen}  In fact, the state of the universe, in this period, is
described by a density matrix $\rho (q,q^{\prime })$ where $q,q^{\prime }\in
Q$. Among all the configuration variables $q$ we may choose a particular
one, as the hand of our clock, let us say the radius of the universe $a$.
Then $\rho (q,q^{\prime })=\rho (a,x,a,x^{\prime }),$ where $x$ symbolizes
the rest of the variables. But to use, $a$ as the hand of our clock, it is
necessary to have decoherence for $a\neq a^{\prime }$, e. g. $\rho
(a,x,a^{\prime },x^{\prime })=0$ if $a\neq a^{\prime }$ or $\rho
(a,x,a^{\prime },x^{\prime })\sim \delta (a-a^{\prime })\rho (a,x,x^{\prime
})$. Namely, at least one variable, e. g. $a,$ must have classical
properties, if not we will be forced to make a theory with two ''times'' $a$
and $a^{\prime }$. But if we have just one $a$ we are not more in full QC,
but in a semiclassical regime , as those of section 4, where in fact, we
will find decoherence of the radius $a$ \cite{Lombardo}. We conclude that,
even if some generalization of time may be defined in full quantum gravity
period, the usual time, with its known properties, cannot exist in such
regime.
\bibitem{Pajaro}  If only spaces $\Phi _{-}$ and $\Phi _{-}^{\times }$ were
defined, for every state $\rho \in \Phi _{-}$ and every observable $A\in
\Phi _{-}^{\times }$ (cf. eq.\ref{T.1.15}) we can compute the ''mean
value'': 
\[
<A>_\rho =A[\rho ]
\]
This number is well defined by the linear operator $A$ acting on the state $%
\rho $. Really this is the maximal answer QC can give: a typical ''mean
value'' for every observable and for every quantum state of the universe,
since the ensemble of QC has just one system: the universe. Therefore
the''mean value'' is not a proper mean value but just a ''typical'' one.

If we want to know the probability $p(q)$ to find the universe at a certain $%
q$ and if $Q$ is the position operator: 
\[
Q=\int q|q><q|dq
\]
and $\rho $ is a pure state $\rho =|\varphi ><\varphi |$, we need to define
a norm and the inner product of ${\cal H}$ to compute: 
\[
p(q)=|<\varphi |q>|^2
\]
But we know that the notion of probability in an ensemble of just one
element in not a reasonable one. So really it is more logical to work just
with the couple $\Phi _{-},\Phi _{-}^{\times }$, and with the typical values 
$A[\rho ]$ than to use the notion of probability. Nevertheless in this paper
we will use the more familiar ''probabilistic approach'' of QC, where we
need a norm like \ref{T.1.2}. We will study the ''typical approach'' of QC,
where the norm is superfluous, in a forthcoming paper.
\bibitem{WdW}  If $\rho =|\Psi ><\Psi |$ is a pure state these equations
coincide with WDW-equation (\ref{T.1.1}).
\bibitem{SE}  As we will see in the model of section 2, we must restore
variable $a$ in $|$ $\omega ,out>,$ which is really a function of $\eta
=\eta (a),$ and then multiply it by an adequate prefactor, to get a solution
of WDW equation. With these modifications $|\omega ,out>$ becomes a vector
of superspace.
\bibitem{Necessary}  We do not know which are the necessary and sufficient
conditions in order to be sure that this structure would exist. Any how
there is an obvious necessary condition: the superspace must be time
oriented, namely two subsuperspaces $\Phi _{-}$ and $\Phi _{+}$ must be
found such that $K:\Phi _{-}\rightarrow \Phi _{+}\neq \Phi _{-}$, being $K$
the Wigner operator (namely the complex conjugation).
\bibitem{Lombardo?}  As $(H_{-}^2)^{*}=H_{+}^2$ then if $\Psi (\omega
)=<\omega ,out|\Psi >\in H_{-}^2$ we have $\Psi ^{*}(\omega )=<\omega
,out|\Psi >^{*}\in H_{+}^2.$ Then we can foresee that the substitution of $%
\Phi _{-}$ by $\Phi _{+}$ will become the time inversion when time will be
defined, namely $K$, the Wigner operator, will become this inversion.
\bibitem{Cont}  Of course, we will have such kind of continuous spectrum,
only if the spatial geometry of the universe is open. In closed models we
can only suppose that the discrete eigenvalues are so closed that they can
be considered as a continuous spectrum in some approximation. This fact, and
the problems that we will face with an expanding-contracting universe, shows
that our formalism is much better adapted to open geometries
\bibitem{DBohm}  Following D. Bohm we can also say that WDW equation is
exactly equivalent to the system: 
\[
[S^{\prime }(a)]^2=2V(a)+\frac i{2M^2}S^{\prime \prime }(a)
\]
\[
iS^{\prime }(a)\frac \partial {\partial a}\chi =h\chi -\frac 1{2M^2}\frac{%
\partial ^2}{\partial a^2}\chi 
\]
where the $M^{-2}$ order corrections can be considered as a gravitational
correction to the Hamilton-Jacobi equation and as a quantum potential that
must be added to the classical potential of hamiltonian $h.$ Theoretically
this system can be exactly solved. See:

Bohm D., Hiley B. J. {\it The undivided universe, }Routledge London, 1993.
\bibitem{S}  Precisely and repeating what we have anticipated in \cite{SE}, $%
|\chi >\in \phi _{-}^{out}$ will be a function of $\eta $ and a functional
of field $\varphi $. But $a=a(\eta )$ so $\chi =\chi (a,\varphi )$ is the
function of eq. (\ref{WKB}), that multiplied by the prefactor $\exp [iMS(a)]$
is $\Psi (a,\varphi ),$ a solution of the WDW equation. So it is $\Psi
(a,\varphi )$ the function that really belongs to $\phi _{-}.$ This fact
proves that the asymmetry exists either in the spaces of $\chi $ or $\Psi $
functions and therefore in the full quantum superspace.
\bibitem{Modifications}  The main difference of this curved space-time
formalism, with the flat space usual one, is that in the former case we have
two unequivalent vacua; $|0,in>$ and $|0,out>,$ while in the latter one we
have just one: $|0>$, such that $K|0>=|0>,$ where $K$ is the Wigner
time-reversal operator. In the curved space time case $K|0,in>\neq |0,out>$
and therefore we have two equations: 
\[
K:\phi _{-}^{out}\rightarrow \phi _{+}^{out}\neq \phi _{-}^{out}
\]
\[
K:\phi _{+}^{in}\rightarrow \phi _{-}^{in}\neq \phi _{+}^{in}
\]
In the usual case the ''in'', ''out'' superscript would be absent and
therefore, there would only be one equation. These differences, with the
usual case, must be taken into account but they are not very important.
\bibitem{A}  Eventually the location of the poles can be such that it turns
out to be impossible to find the curve $\Gamma .$ Then we can use a set of
curves \{$\Gamma _n\}$, such that $\Gamma _n$ goes below the poles $\Omega
_1,\Omega _2,...\Omega _n,$ and take $n\rightarrow \infty $ at the end of
calculations
\bibitem{Baker}  This fact is evident if we consider a Baker's
transformation \cite{Mackey}: Let us consider a distribution function with
compact support, and such that it belongs to ${\cal L}=L^2([0,1]\times
[0,1]).$ In the far future this support becomes a set of horizontal strips,
and in the limit a set of horizontal straight lines, such that it is dense
in $[0,1]\times [0,1].$ No function of ${\cal L}$ has such a support and
therefore the equilibrium state does not belong to ${\cal L}$.
\end{thebibliography}
\end{document}